\documentclass[prc,nofootinbib,showpacs]{revtex4}

\usepackage{graphicx}
\usepackage[usenames,dvipsnames]{color}
\usepackage{amsmath,amssymb,bbold}
\usepackage{hyperref}

\begin{document}

\title{Thermoelectric effect and Seebeck coefficient for hot and dense hadronic matter}

\author{Jitesh R. Bhatt}
\email{jeet@prl.res.in}
\author{Arpan Das}
\email{arpan@prl.res.in}
\author{Hiranmaya Mishra}
\email{hm@prl.res.in}
\affiliation{Theory Division, Physical Research Laboratory,
Navrangpura, Ahmedabad 380009, India}

\begin{abstract}
We investigate the thermoelectric effect for baryon rich plasma produced in heavy ion collision experiments. We estimate the associated 
Seebeck coefficient
for the hadronic matter. Using kinetic theory within relaxation time approximation we calculate the 
Seebeck coefficient of a hadronic medium with a temperature gradient. The calculation is performed for hadronic matter modeled by the hadron
resonance gas model with hadrons and resonance states up to a cutoff in the mass as 2.25 GeV. 
We argue that the thermoelectric current produced by such effect can produce a magnetic
field in heavy ion collision experiments.
\end{abstract}

\maketitle

\section{INTRODUCTION}

\label{intro}

 Transport coefficients of strongly interacting matter under extreme conditions of temperature, density and/or magnetic fields
have been one of the most challenging interests in the field of strong interaction physics.
In the context of relativistic heavy ion collision
experiments (RHIC), these are important input parameters that enter in the dissipative relativistic hydrodynamics as well as
transport simulations, that are being used to describe the evolution of the matter subsequent to a heavy ion collision.
Indeed, a small shear 
viscosity to entropy ratio ($\eta/s$) was necessary to explain the flow data \cite{HeinzSnellings2013,RomatschkeRomatschke,KSS}.
The smallness of $\eta/s$ is significant
in connection with the conjectured lower bound of $\eta/s=1/4\pi$ which initiated a flurry of activity in understanding 
this from microscopic  theory \cite{KSS}. The other viscosity coefficient, the bulk viscosity $\zeta$, was later realized to be also important 
in the dissipative hydrodynamics describing the QGP evolution \cite{DobadoTorres2012,sasakiRedlich2009,sasakiRedlich2010,KarschKharzeevTuchin2008,
FinazzoRougemont2015,WiranataPrakash2009,JeonYaffe1996}. The bulk viscosity scales as the conformal measure $(\epsilon-3P)/T^4$,
and becomes very large near the quark hadron phase transition as inferred from lattice QCD simulations. The effect of large bulk viscosity
has been investigated for particle spectra and flow coefficients. Effects of bulk viscosity and shear viscosity on the elliptic flow
have also been investigated.
Both $\eta/s$ and $\zeta/s$ as a function of temperature show nonmonotonic behavior near the  critical temperature $T_c$ \cite{DobadoTorres2012,sasakiRedlich2009,sasakiRedlich2010,KarschKharzeevTuchin2008,
FinazzoRougemont2015,WiranataPrakash2009,JeonYaffe1996}. 
 In case of noncentral heavy ion collision, a large magnetic field is also expected to be produced.
The strong magnetic field so produced has exciting possibilities of observing different CP violating effects like chiral magnetic effect
and chiral vortical effect  \cite{kharzeevbook}. The effect of such strong magnetic field on enhancing the elliptic flow coefficient has been investigated.
 Such phenomenologically interesting manifestation of strong magnetic fields also requires that a strong magnetic field survives for at least few
Fermi proper time. The crucial parameter that enters in the  time dependence of the magnetic field in the medium is the electrical 
conductivity, 
$\sigma_{el}$\cite{TuchinMHD,MHD1,MHDajit,MoritzGreif,electricalcond1,electricalcond2,electricalcond3,electricalcond4,
electricalcond5,electricalcond6,electricalcond7,electricalcond8,electricalcond9,electricalcond10,electricalcond11,electricalcond12,
electricalcond13,electricalcond14,electricalcond15}. These transport coefficients for quark matter can be 
estimated using different approaches like perturbative QCD, and different effective models  (see \cite{danicol2018,PrakashVenu,WiranataPrakash2012,
KapustaChakraborty2011,Toneev2010,Plumari2012,Gorenstein2008,Greiner2012,TiwariSrivastava2012,GhoshMajumder2013,Weise2015,GhoshSarkar2014,
WiranataKoch,WiranataPrakashChakrabarty2012,Wahba2010,Greiner2009,KadamHM2015,Kadam2015,Ghoshijmp2014,Demir2014,Ghosh2014,smash,bamps,bamps2,
urqmd1,GURUHM2015}).
The other transport coefficient that has been important for heavy ion collisions involving high baryon densities, 
is the thermal conductivity. The effects of thermal conductivity on relativistic hydrodynamics have only recently been studied
\cite{danicol2014,Kapusta2012}.

 In the present work, we investigate another related coefficient  relevant for high density heavy ion collision, namely, the thermoelectric behavior of the strongly interacting matter in heavy-ion collisions.
  The phenomenon in which
 a temperature gradient  in a conducting material is  converted to electrical current and vice versa is known as thermoelectric effect,
which is  also
known as the Seebeck effect. The Seebeck effect in a conductor is a  manifestation of the fact that when there exists a temperature gradient,
 the charge carriers would diffuse toward the 
 region of lower temperature. This diffusion continues till the electric field generated by the motion of charge
carriers becomes strong enough to stop this motion.
  The  Seebeck  coefficient is defined as  the  electric field  produced  in  a  conducting medium  due to a temperature  gradient
 when  the  electrical  current  is set to  zero \cite{callen,sofo}.
 Such properties have been investigated in various condensed matter systems
quite extensively. 
This includes the study of the Seebeck effect in superconductors \cite{conseeb1,conseeb2}, the Seebeck effect in the graphene-superconductor junction
\cite{conseeb3}, thermoelectric signatures of a Majorana bound state coupled to a quantum dots \cite{conseeb4}, thermoelectric 
properties of high temperature cuprates\cite{conseeb5}, thermoelectric properties of superconductor-ferromagnetic tunnel junctions
\cite{conseeb6}, Seebeck coefficient in low dimensional organic metals \cite{conseeb7}etc.

  In  the present investigation, we study the Seebeck effect for
 hot and dense hadronic matter. It may be noted that in the usual condensed matter systems, the thermoelectric effect requires 
 only a temperature gradient, as the ions in the lattice are stationary. On the other hand, e.g., in 
 an electron-positron plasma, just having a temperature
 gradient is not enough to lead to any thermoelectric current. This will be similar in quark gluon plasma (QGP) with 
zero baryon density. However,  the situation is different at finite
 baryon chemical potential, when the number of baryons and antibaryons are different. In the presence of a temperature gradient,
 there will be net thermoelectric current driven by the temperature gradient as there will be unequal number of positive and negative
charge carriers.
 For the heavy-ion collisions at Facility for Antiproton and Ion Research (FAIR) at Darmstadt \cite{fairref} and in 
Nuclotron-based Ion Collider fAcility (NICA) at Dubna \cite{nica} one  expects a baryon-rich strongly
 interacting medium being created. In these cases, the thermalization 
of the strongly interacting medium is expected which is not electrically charge neutral. 
The strongly interacting matter created in heavy-ion collisions can have large temperature gradient between the 
 central and the peripheral regions of the collision. Thus, by allowing the possibility of a
temperature gradient, one can argue that the electric current in the medium will not only depend on the given electric
field but also on the temperature gradient. 
Keeping the above motivation in mind, we calculate the Seebeck coefficient of hadron resonance gas within the kinetic theory
framework in  the relaxation time approximation. As the mesons carry no baryonic chemical potential, one might naively expect that the mesons
(dominantly pions) will not contribute to the Seebeck coefficient. However, as we shall see, mesons also become relevant for
 the total Seeback coefficient of such hot  and dense hadronic matter.

 The hadronic phase of the strongly interacting medium created in heavy ion collisions are well
described in terms of the hadron resonance gas (HRG) model, at chemical freeze-out \cite{HRG1,HRG2}. 
If one assumes strange and nonstrange particles
freeze out in the same manner then HRG model has only two parameters $T$ and $\mu$ in its simplest form, where $T$ and $\mu$ are temperature 
and baryon chemical potential respectively. HRG model has been very successful in explaining the experimental result 
of the thermal abundance of different particle ratios in the heavy ion collisions, for a given temperature and baryon chemical potential
\cite{HRG3}. Naively one expects that 
a system of hadrons will be an interacting system and in general, thermodynamics of interacting hadrons can be nontrivial. However, 
it has been shown that 
 in the presence of narrow resonances, the thermodynamics of interacting gas of hadrons can be approximated by the non-interacting
gas of hadrons and resonances \cite{HRG4,HRG5}. Due to its simple structure and minimal parameters, HRG model has been well explored regarding
thermodynamics \cite{thermodynamicsHRG1,thermodynamicsHRG2}, conserved charge fluctuations\cite{hrgfluc3,hrgfluc4,hrgfluc5,hrgfluc6,
hrgfluc7} as well as transport coefficients for
hadronic matter\cite{MoritzGreif,PrakashVenu,WiranataPrakash2012,
KapustaChakraborty2011,Toneev2010,Plumari2012,Gorenstein2008,Greiner2012,TiwariSrivastava2012,GhoshMajumder2013,Weise2015,GhoshSarkar2014,
WiranataKoch,WiranataPrakashChakrabarty2012,Wahba2010,Greiner2009,KadamHM2015,Kadam2015,Ghoshijmp2014,Demir2014,Ghosh2014,smash,
bamps,bamps2,electricalcond2,electricalcond3,Plumari2012,urqmd1,GURUHM2015}. One 
can improve upon the ideal HRG model e.g. including excluded volume HRG model
\cite{stockerRischke,GURUHM2015}. 
In present investigation, we will only discuss within the ideal hadron
resonance gas model at finite temperature (T) and baryon chemical potential ($\mu$) to estimate the Seebeck coefficient.
We would like to mention here that, although, thermoelectric effect has been studied extensively in the condensed matter systems, 
it has not been studied in the context of heavy ion collisions. The present work is a first step in that direction limited to hot and dense
hadronic matter.

This paper is organized as follows, in Sec.\ref{formalism} we introduce the formalism of the Seebeck coefficient from kinetic theory within relaxation
time approximation. We also generalize it to a multicomponent system. In section Sec.\ref{HRGmodel} we briefly discuss the HRG model and estimate the relaxation time within the same model.
In Sec. \ref{results} we present and discuss the results for the Seebeck coefficient. Finally we summarize our work with an outlook in the
conclusion section.

\section{Boltzmann equation in relaxation time approximation and Seebeck coefficient for multicomponent system}
\label{formalism}

 We consider here the linearized Boltzmann equation in relaxation time approximation.
 For a linear problem or weak external fields, the Boltzmann equation can be interpreted as a linear expansion of the distribution 
function around the equilibrium distribution function, hence $f(\vec{k})$ which denotes out of equilibrium distribution function, is not very far
from equilibrium. Due to strong interaction equilibrium is achieved locally and electromagnetic force will take the system out of equilibrium.
Using the linear response approximation we write the Boltzmann equation as \cite{linearizedBoltzmann},

\begin{equation}
 \vec{v}.\vec{\nabla}_{\vec{r}}f_0+\vec{F}.\vec{\nabla}_{\vec{k}}f_0 = -\frac{f(\vec{r},\vec{k})-f_0(\vec{r},\vec{k})}{\tau(\vec{k})}
 \equiv -\frac{f^{(1)}(\vec{r},\vec{k})}{\tau(\vec{k})} ,
 \label{eq1}
\end{equation}
where, $f_0$ denotes equilibrium distribution function, $f$ denotes out of equilibrium distribution function, and $\tau$ denotes the relaxation 
time of the system.  The local equilibrium distribution function is considered to be of the following form \cite{linearizedBoltzmann}:
\begin{equation}
 f_0(\vec{r},\vec{k})=\frac{1}{1+\exp(\frac{E(\vec{k})-\mu(\vec{r})}{T(\vec{r})})}.
 \label{eq2}
\end{equation}
where $E$, $T$, and $\mu$ denote energy, temperature, and chemical potential respectively, and they are 
functions of position vector $\vec{r}$.
 Relaxation time encodes interaction processes of the microscopic 
theory and these interaction does not depend on the coordinate position. It is further assumed that cross section of
the local interaction is independent of spatial coordinates.
\noindent
 Using Eq.\eqref{eq2} spatial gradient of distribution function can be written as,
\begin{equation}
 \vec{\nabla}_{\vec{r}}f_0(\vec{r},\vec{k}) =-\frac{f_0(\vec{r},\vec{k})(f_0(\vec{r},\vec{k})-1)}{T}
\bigg(\vec{\nabla}_{\vec{r}}
 \mu(\vec{r})+\left(E(\vec{k})-\mu(\vec{r})\right)\vec{\nabla}_{\vec{r}}\ln T(\vec{r})\bigg).
 \label{eq3}
\end{equation}
To get the above equation we have used,
\begin{equation}
 \frac{\partial f_0(\vec{r},\vec{k})}{\partial E(\vec{k})}=
 \frac{f_0(\vec{r},\vec{k})(f_0(\vec{r},\vec{k})-1)}{T}
  \label{eq4}
\end{equation}
Similarly for the second term in the L.H.S of Eq.\eqref{eq1}, momentum derivative of distribution function is given by,

\begin{equation}
 \vec{\nabla}_{\vec{k}}f_0(\vec{r},\vec{k})=\frac{\partial f_0(\vec{r},\vec{k})}{\partial E(\vec{k})} \vec{\nabla}_{\vec{k}}E(\vec{k})=
 \frac{f_0(\vec{r},\vec{k})(f_0(\vec{r},\vec{k})-1)}{T}\vec{v},
 \label{eq5}
\end{equation}
where we have written, $\vec{\nabla}_{\vec{k}}E(\vec{k}) =  \vec{v}$.

Using Eq.\eqref{eq3} and Eq.\eqref{eq5} the Boltzmann equation (Eq.\eqref{eq1}) can be recasted as, with the force 
$\vec{F}=e\vec{\mathcal{E}}$,

\begin{eqnarray}
 \frac{f^{(1)}(\vec{r},\vec{k})}{\tau(\vec{k})} &=& -\frac{f_0(\vec{r},\vec{k})(f_0(\vec{r},\vec{k})-1)}{T} \vec{v}.
 \bigg(e\vec{\mathcal{E}}-T(\vec{r})\vec{\nabla}_{\vec{r}}\bigg(\frac{\mu(\vec{r})}{T(\vec{r})}\bigg)-\frac{E(\vec{k})}{T(\vec{r})}
 \vec{\nabla}_{\vec{r}}T(\vec{r})\bigg).
 \label{eq7}
\end{eqnarray}
 From now on we omit the explicit functional dependence of distribution function ($f_0$),
 chemical potential ($\mu$), temperature ($T$) and relaxation time ($\tau$) unless otherwise stated.

 Electric current density is defined as,
 \begin{equation}
  \vec{j}=\frac{e g}{(2\pi)^3}\int_{-\infty}^{\infty}\vec{v}f d^3\vec{k}=\frac{e g}{(2\pi)^3}\int_{-\infty}^{\infty}\vec{v}f^{(1)} d^3\vec{k},
\label{eq8}
  \end{equation}
here $g$ is the degeneracy factor. In writing the second step we have used the fact that the equilibrium distribution function, being isotropic does not
leads to a current. Similarly the heat current is defined as for example see Eq. 2.42 of \cite{thermoelectrics},

\begin{equation}
 \vec{j}_Q=\frac{ g}{(2\pi)^3}\int_{-\infty}^{\infty}(E-\mu)\vec{v}f^{(1)} d^3\vec{k}.
\label{eq9}
 \end{equation}
Note that heat current here is denoted as $ \vec{j}_Q$ (``$Q$'' stands for heat) and $\vec{j}$ denotes the electric current throughout this paper.
Using Eq.\eqref{eq7} in Eq.\eqref{eq8} and Eq.\eqref{eq9}  for $f^{(1)}$ electric and the heat current respectively given as,

\begin{equation}
 \vec{j}=-\frac{e g}{(2\pi)^3}\int_{-\infty}^{\infty}\frac{f_0(f_0-1)}{T}\tau\vec{v}\bigg[\vec{v}.\bigg(e\vec{\mathcal{E}}-
 T\vec{\nabla}\left(\frac{\mu}{T}\right)-\frac{E}{T}\vec{\nabla} T\bigg)\bigg]d^3\vec{k}.
\end{equation}

\begin{equation}
 \vec{j}_Q=-\frac{ g}{(2\pi)^3}\int_{-\infty}^{\infty}(E-\mu)\frac{f_0(f_0-1)}{T}\tau\vec{v}\bigg[\vec{v}.\bigg(e\vec{\mathcal{E}}-
 T\vec{\nabla}\left(\frac{\mu}{T}\right)-\frac{E}{T}\vec{\nabla} T\bigg)\bigg]d^3\vec{k}.
\end{equation}
For an isotropic medium $\vec{j},\vec{j}_Q$ reduced to,

\begin{equation}
 \vec{j}=-\frac{e g}{(2\pi)^3}\int_{-\infty}^{\infty}\frac{f_0(f_0-1)}{T}\tau \frac{v^2}{3}\bigg[\bigg(e\vec{\mathcal{E}}-
 T\vec{\nabla}\left(\frac{\mu}{T}\right)-\frac{E}{T}\vec{\nabla} T\bigg)\bigg]d^3\vec{k},
\label{eq12}
 \end{equation}

\begin{equation}
 \vec{j}_Q=-\frac{ g}{(2\pi)^3}\int_{-\infty}^{\infty}(E-\mu)\frac{f_0(f_0-1)}{T}\tau \frac{v^2}{3}\bigg[\bigg(e\vec{\mathcal{E}}-
 T\vec{\nabla}\left(\frac{\mu}{T}\right)-\frac{E}{T}\vec{\nabla} T\bigg)\bigg]d^3\vec{k}.
\label{eq13}
 \end{equation}

Note that due to the presence of the external force, the velocity will not be isotropic in general.
But since the external force is considered to be small, the change in the velocity can be ignored. For later calculations, it is convenient to rewrite 
the momentum integration in Eq.\eqref{eq12}, Eq.\eqref{eq13} in terms of integration over energies so that,

\begin{equation}
 \vec{j}=-\frac{2 e}{3 m}\int_{0}^{\infty}\frac{f_0(f_0-1)}{T}\tau ED(E)\bigg[\bigg(e\vec{\mathcal{E}}-
 T\vec{\nabla}\left(\frac{\mu}{T}\right)-\frac{E}{T}\vec{\nabla} T\bigg)\bigg]dE,
\label{eq16}
 \end{equation}

\begin{equation}
 \vec{j}_Q=-\frac{2 }{3 m}\int_{0}^{\infty}(E-\mu)\frac{f_0(f_0-1)}{T}\tau ED(E)\bigg[\bigg(e\vec{\mathcal{E}}-
 T\vec{\nabla}\left(\frac{\mu}{T}\right)-\frac{E}{T}\vec{\nabla} T\bigg)\bigg]dE,
\label{eq17}
 \end{equation}
 where density of states is defined as, $D(E)dE\equiv\frac{d^3k}{4\pi^3}$ and for particles in nonrelativistic limit $v^2=\frac{2E}{m}$.
 The expressions for $\vec{j}$ and $\vec{j}_Q$ can be written in a compact manner, by defining the integral $\mathcal{L}_{ij}$ as,
if we define the following integral,

\begin{equation}
 \mathcal{L}_{ij}=-\frac{2}{3m}\int \frac{f_0(f_0-1)}{T}E^i\tau^jD(E)dE,
\label{eq18}
 \end{equation}
where $i,j$ are not tensor indices, rather they denote the number of times $E$ and $\tau$ appear in the expression. In terms of $\mathcal{L}_{ij}$
one writes expressions for the currents as,

\begin{equation}
 \vec{j}= e\mathcal{L}_{11}\bigg(e\vec{\mathcal{E}}-T\vec{\nabla}\left(\frac{\mu}{T}\right)\bigg)-e\mathcal{L}_{21}\frac{\vec{\nabla}T}{T},
 \label{eq19}
\end{equation}

\begin{equation}
 \vec{j}_Q= \bigg(\mathcal{L}_{21}-\mu\mathcal{L}_{11}\bigg)\bigg(e\vec{\mathcal{E}}-T\vec{\nabla}\left(\frac{\mu}{T}\right)\bigg)
 -\bigg(\mathcal{L}_{31}-\mu\mathcal{L}_{21}\bigg)\frac{\vec{\nabla}T}{T}.
 \label{eq20}
\end{equation}
One can further assume that chemical potential has no spatial dependence. In this approximation $\vec{j}$ and $\vec{j}_Q$ becomes,

\begin{equation}
 \vec{j}=e^2\mathcal{L}_{11}\vec{\mathcal{E}}-(e^2\mathcal{L}_{11})\frac{\left(\mathcal{L}_{21}-\mu\mathcal{L}_{11}\right)}{e\mathcal{L}_{11}T}\vec{\nabla}T,
\label{eq21}
 \end{equation}
 
 \begin{equation}
  \vec{j}_Q=\bigg(\mathcal{L}_{21}-\mu\mathcal{L}_{11}\bigg)e\vec{\mathcal{E}}-
  \bigg(\mathcal{L}_{31}-\mu\mathcal{L}_{21}+\mu^2\mathcal{L}_{11}\bigg)\frac{\vec{\nabla}T}{T}.
  \label{eq22}
 \end{equation}

The Seebeck coefficient $S$ is determined by setting $\vec{j}=0$, so that the electric field gets related with the temperature gradient 
as \cite{thermoelectrics},

\begin{equation}
\vec{\mathcal{E}} = \left(\frac{\mathcal{L}_{21}-\mu\mathcal{L}_{11}}{e\mathcal{L}_{11}T}\right)\vec{\nabla}T\equiv S\vec{\nabla}T ,
\label{eq23}
\end{equation}

From Eq.\eqref{eq21} electrical conductivity can be identified as,

\begin{equation}
 \sigma_{el} = e^2\mathcal{L}_{11}.
 \label{eq25}
\end{equation}
Hence the electric current can be expressed in terms  of $\sigma_{el}$ and $S$ as \cite{sofo},

\begin{equation}
 \vec{j}=\sigma_{el}\vec{\mathcal{E}}-\sigma_{el}S \vec{\nabla}T.
 \label{eq26}
\end{equation}
In a similar way, when $\vec{\nabla}\mu =0$, the heat current can be expressed as \cite{sofo},

\begin{equation}
 \vec{j}_Q=T\sigma_{el}S\vec{\mathcal{E}}-k_0\vec{\nabla}T,
 \label{eq27}
\end{equation}
where, 

\begin{equation}
 k_0=\frac{1}{T}\left(\mathcal{L}_{31}-2\mu\mathcal{L}_{21}+\mu^2\mathcal{L}_{11}\right).
 \label{eq28}
\end{equation}
Using Eq.\eqref{eq26} and Eq.\eqref{eq27}, heat current $\vec{j}_Q$ can be expressed in terms  of electric current $\vec{j}$,

\begin{equation}
 \vec{j}_Q=TS\vec{j}-\left(k_0-T\sigma_{el}S^2\right)\vec{\nabla}T.
 \label{eq29}
\end{equation}
From Eq.\eqref{eq29} we can identify the Peltier coefficient and thermal conductivity respectively,

\begin{equation}
 \Pi=TS,
\end{equation}

\begin{equation}
 k = k_0-T\sigma_{el}S^2.
\end{equation}
Seebeck coefficient as given in Eq.\eqref{eq23} is a standard result for condensed matter systems \cite{thermoelectrics} and it is 
obtained by considering a single species of charged particle. However for the case of heavy-ion collisions there can be multiple charged particle
species and we need to generalize above result. The total Seebeck coefficient of the system can not be given by Eq.\eqref{eq23}.
For multiple species case, the total electric current is a vector sum of the currents due to different species and thus one writes:

\begin{equation}
 \vec{j}= \vec{j}_{(1)}+\vec{j}_{(2)}+\vec{j}_{(3)}+......=\sum_i  \vec{j}_{(i)},
\end{equation}
with $\vec{j}_{(i)}$ being electrical current for the $i$'th species,
\begin{equation}
 \vec{j}_{(i)}=e^2_{(i)}\mathcal{L}_{11}^{(i)}\vec{\mathcal{E}}-\frac{e_{(i)}}{T}\left(\mathcal{L}_{21}^{(i)}-\mu\mathcal{L}_{11}^{(i)}\right).
\end{equation}
Hence,

\begin{equation}
 \vec{j}=\left(e^2_{(1)}\mathcal{L}_{11}^{(1)}+e^2_{(2)}\mathcal{L}_{11}^{(2)}+......\right)\vec{\mathcal{E}}
 -\left(\frac{e_{(1)}}{T}\left(\mathcal{L}_{21}^{(1)}-\mu\mathcal{L}_{11}^{(1)}\right)+\frac{e_{(2)}}{T}\left(\mathcal{L}_{21}^{(2)}-
 \mu\mathcal{L}_{11}^{(2)}\right)+....\right)\vec{\nabla}T.
\end{equation}
The Seebeck coefficient of the multispecies system can now be defined as,

\begin{align}
 S=\frac{\sum_i\frac{e_{(i)}}{T}(\mathcal{L}_{21}^{(i)}-\mu\mathcal{L}_{11}^{(i)})}{\sum_i e_{(i)}^2\mathcal{L}_{11}^{(i)}}\equiv 
  \frac{\sum_i S^{(i)}e^2_{(i)}\mathcal{L}_{11}^{(i)}}{\sum_i e_{(i)}^2\mathcal{L}_{11}^{(i)}},
  \label{eq38}
\end{align}
where, we have defined the Seebeck coefficient of each species as
\begin{equation}
S^{(i)}=\left(\mathcal{L}_{21}^{(i)}-\mu\mathcal{L}_{11}^{(i)}\right)/e_{(i)}\mathcal{L}_{11}^{(i)}T.
\label{si}
\end{equation}
Thus the total Seebeck coefficient $S$  of the medium is a weighted average of the Seebeck coefficients of the 
individual species.

In this context, it may be relevant to note that, similar to the expression for 
electrical conductivity,   the Seebeck coefficient in  Drude picture has been estimated in Ref.\cite{fujita}.
The Seebeck coefficient for a single species in the Drude picture can be written as
$S=\frac{2\ln2}{d}(1/qn)\epsilon_Fk_B\frac{\mathcal{N}_0}{V}$, where $q$ is the charge of the 
particle,$d$, being the spatial dimensionality of the system,
$n$ is the number density,
$\epsilon_F$ is the Fermi energy, $k_B$ is the Boltzmann constant, $\mathcal{N}_0$ is the density of the states at $\epsilon_F$ and V is the 
volume. The kinetic theory expression for single species on the other hand is given by Eq.\eqref{si}. Let us further note that while the
individual Seebeck coefficient is independent of the average relaxation time, which cancels from the numerator and 
denominator of Eq.\eqref{si}, the Seebeck coefficient of the multicomponent system is dependent on the relaxation time 
for each species as may be clear from Eq.\eqref{eq38}.
 Thus estimating the 
Seebeck coefficient of a system of charged particles reduces to calculating the Seebeck coefficient of each species
using Eq.\eqref{si} and the different ${\cal L}_{ij}$ given in Eq.\eqref{eq18}.
In the following section we estimate the Seebeck coefficient of the multicomponent  hadronic system within the HRG model.

\section{HADRON RESONANCE GAS MODEL}
\label{HRGmodel}
The central quantity in hadron resonance gas model is the thermodynamic potential which is given by \cite{KadamHM2015},

\begin{equation}
 \log Z(\beta,\mu,V)=\int dm (\rho_M(m)\log Z_b(m,V,\beta,\mu)+\rho_B(m)\log Z_f(m,V,\beta,\mu)),
 \label{eq43}
\end{equation}
where, the gas of noninteracting pointlike hadrons and their resonances is contained in the volume $V$ at a temperature $T = 1/\beta$ and baryon chemical potential
$\mu$. $Z_b$ and $Z_f$ corresponds to the partition functions of free bosons (mesons) and fermions (baryons) respectively with mass $m$. 
Further, $\rho_B$ and $\rho_m$ are the spectral functions of free bosons (mesons) and fermions (baryons) respectively. The spectral
densities  encode the hadron properties. Various thermodynamic quantities can be calculated from the logarithm of the partition function as given in
Eq.\eqref{eq43} by taking derivatives with respect to the thermodynamic parameters T, $\mu$ and the volume $V$, once the spectral density is specified. One common approach in HRG models is in taking all the  hadrons and their resonances below
a certain mass cutoff $\Lambda$ to estimate the thermodynamic potential. This is achieved by taking the spectral density
$\rho_{B/M}(m)$ as, 

\begin{equation}
 \rho_{B/M}(m)=\sum_i^{M_i<\Lambda}g_i\delta(m-M_i),
 \label{eq44}
\end{equation}
where the sum is taken over all the hadron and resonance states up to a mass that are less than the cutoff $\Lambda$.
 In  Eq.\eqref{eq44}, $M_i$ are
the masses of the known hadrons and their resonances and $g_i$ is the corresponding degeneracy, which includes spin and isospin quantum numbers. Although 
in this work we have used the discrete spectrum, it is important to mention that HRG model including discrete particle spectrum can explain
lattice QCD data for trace anomaly up to temperature $\sim 130 $ MeV \cite{HRGMuller}. Including Hagedron spectrum along with the discrete
spectrum for the spectral function can explain lattice QCD data for QCD trace anomaly up to $T \sim$ 160 MeV \cite{HRGMuller}. Once the partition function of the 
HRG model is known from Eq.\eqref{eq43}, thermodynamic quantities like pressure, energy density, number density etc. can be calculated 
using standard thermodynamic relations. For details of thermodynamics 
of HRG model, see e.g. Ref.\cite{HRG1}. In terms of discrete spectral function Eq.\eqref{eq44}, the integrals
$\mathcal{L}_{11}$ and $\mathcal{L}_{21}$ for each species as in Eq.\eqref{eq18} in Boltzmann approximation can be expressed as,

\begin{equation}
 \mathcal{L}_{11}^i=\frac{\tau_i g_i}{6\pi^2 T}\int_0^{\infty}\frac{k^4}{k^2+m_i^2}\exp\left(-\frac{(\sqrt{k^2+m_i^2}-\mu B^i)}{T}\right)dk,
 \label{eq45}
\end{equation}
and,
\begin{equation}
 \mathcal{L}_{21}^i=\frac{\tau_i g_i}{6\pi^2 T}\int_0^{\infty}\frac{k^4}{\sqrt{k^2+m_i^2}}\exp\left(-\frac{(\sqrt{k^2+m_i^2}-\mu B^i)}{T}\right)dk,
\label{eq46}
 \end{equation}
 where $B^i$ is the baryon number of the $i-$th species.  Let us note that the Boltzmann approximation for baryons
 is a reasonable approximation as long as $m_{nucleon}-\mu\geq T$ \cite{thermodynamicsHRG1}.
In the present work, we have taken all the hadrons and their resonances with masses up to the cut off $\Lambda \sim 2.25$GeV.
Specifically, we for baryons, the maximum mass is up to 2.252 GeV while for mesons the maximum mass is up to 2.011 GeV.
We considered hadrons and resonances for which $m_i\leq \Lambda\simeq 2.25$ GeV. \\

Let us note that, while the individual Seebeck coefficient is independent of the relaxation time , the Seebeck coefficient of the medium is not. In the following we
therefore estimate the same.
The relaxation time is defined as \cite{lataguruhm,KapustaChakraborty2011},

\begin{align}
\tau_a^{-1}(E_a)=\sum_{bcd}\int\frac{d^3p_b}{(2\pi)^3}\frac{d^3p_c}{(2\pi)^3}\frac{d^3p_d}{(2\pi)^3}W(a,b\rightarrow c,d)f_b^0
\end{align}

where the transition rate $W(a,b\rightarrow c,d)$ is,

\begin{align}
W(a,b\rightarrow c,d)=\frac{(2\pi)^4\delta(p_a+p_b-p_c-p_d)}{2E_a2E_b2E_c2E_d}|\mathcal{M}|^2,
\end{align}

$\mathcal{M}$ is the transition amplitude. Then the relaxation time in the center of mass frame can be simplified as, 

\begin{align}
\tau_a^{-1}(E_a)=\sum_b\int\frac{d^3p_b}{(2\pi)^3}\sigma_{ab}v_{ab}f_b^0,
\end{align}

here $\sigma_{ab}$ is the total scattering cross section for the process, $a(p_a)+b(p_b)\rightarrow a(p_c)+b(p_d)$ 
and $v_{ab}$ is the relativistic relative velocity.

\begin{align}
 v_{ab}=\frac{\sqrt{(p_a.p_b)^2-m_a^2m_b^2}}{E_aE_b}
\end{align}

Thus the thermal average relaxation time can be expressed by averaging the relaxation time over $f_a^0$,

\begin{align}
 \tau_a^{-1}=\frac{\int f_a^0\tau^{-1}_a(E_a)dE_a}{\int f^0_adE_a}
\end{align}

The energy averaged relaxation time
$(\tau_a)$, assuming hard sphere scattering can be estimated as \cite{GURUHM2015},

\begin{equation}
 \tau_a^{-1}=\sum_b n_b\langle\sigma_{ab}v_{ab}\rangle,
\end{equation}
where $n_a$ and $\langle\sigma_{ab}v_{ab}\rangle$ represents number density and thermal averaged cross section respectively. The thermal averaged 
cross section for the scattering process $a(p_a)+b(p_b)\rightarrow a(p_c)+b(p_d)$ is given as, assuming hard sphere scattering \cite{gondologelmini},

\begin{equation}
 \langle\sigma_{ab}v_{ab}\rangle = \frac{\sigma}{8Tm_a^2m_b^2K_2(m_a/T)K_2(m_b/T)}\int_{(m_a+m_b)^2}^{\infty}ds\times \frac{[s-(m_a-m_b)^2]}{\sqrt{s}}
 \times [s-(m_a+m_b)^2]K_1(\sqrt{s}/T),
\end{equation}
where $\sigma = 4\pi r_h^2$ is the total scattering cross section for the hard sphere.

\section{RESULTS and DISCUSSIONS}

\label{results}
\begin{figure}[!htp]
\begin{center}
\includegraphics[width=0.48\textwidth]{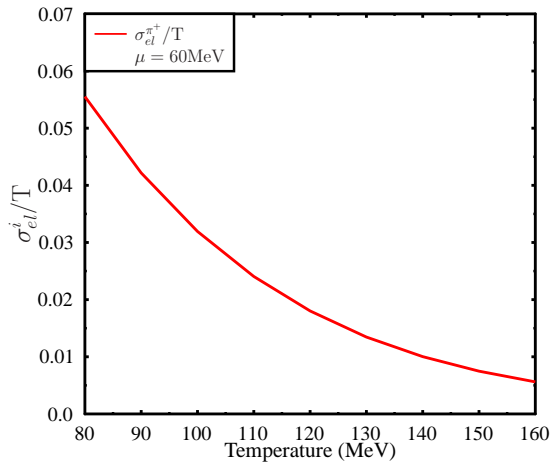}
\caption{Variation of the electrical conductivity of $\pi^+$ with temperature at $\mu=60$ MeV. Order of magnitude estimation
of electrical conductivity of $\pi^+$ is similar to that obtained in Ref.\cite{MoritzGreif}.}
\label{pionsigmael}
\end{center}
\end{figure}

As mentioned earlier for the hadron resonance gas model we shall include all the hadrons and resonances up to a mass cutoff $\Lambda= 2.25$ GeV and include
all the mesons and baryons listed in Ref.\cite{pdg} (also see Appendix A of Ref.\cite{kapustaAlbr} for detailed list of 
particles). The other parameter is the radii of the hard spheres.
We have chosen an uniform radius of
$r_h=0.3$ fm for all the mesons and baryons\cite{GURUHM2015,hmgururadius1}.
We have estimated the Seebeck coefficient of each species using Eq.\eqref{eq23}, Eq.\eqref{eq45} and Eq.\eqref{eq46} for baryon chemical 
potential $\mu=$60 MeV, 80MeV, 100MeV and 150 MeV. For each value of chemical potential we have varied temperature from 80MeV to 160 MeV. Knowing the 
Seebeck coefficient, $\mathcal{L}_{11}$ and $\mathcal{L}_{21}$ of each species, the Seebeck coefficient of the entire system can be found from Eq.\eqref{eq38}. Before doing so, to get a feeling for $\mathcal{L}_{11}$,
let us note that $\mathcal{L}_{11}$ is related to the electrical conductivity as $\sigma_{el}=e^2\mathcal{L}_{11}$. In Fig.\eqref{pionsigmael}
we have plotted $\sigma_{el}$ for $\pi^+$ as a function of temperature. The behavior is similar to as obtained in Ref.\cite{MoritzGreif}.
In fact in the limit of vanishing pion mass $\sigma_{el}\simeq \frac{1}{3T}e^2n_{\pi}\tau_{\pi}$ as in Ref.\cite{MoritzGreif}. 

Next we discuss the Seebeck coefficient for a single species as given by Eq.\eqref{eq23} which can be written as,

\begin{align}
 S^i=\frac{1}{e}\frac{I_{21}}{I_{11}}-\frac{\mu B^i}{eT},
\label{ssingle}
\end{align}
 where the integral $I_{21}$ and $I_{11}$ are given by,
 
 \begin{align}
  I_{11}=\int_{0}^{\infty}\frac{\hat{k}^4}{\hat{k}^2+\hat{m}^2}\exp(-(\sqrt{\hat{k}^2+\hat{m}^2}-\hat{\mu} B^i))d\hat{k},
 \end{align}
and,

\begin{align}
  I_{21}=\int_{0}^{\infty}\frac{\hat{k}^4}{\sqrt{\hat{k}^2+\hat{m}^2}}\exp(-(\sqrt{\hat{k}^2+\hat{m}^2}-\hat{\mu}B^i))d\hat{k},
 \end{align}
 
 where, $\hat{k}=k/T$, $\hat{m}=m/T$ and $\hat{\mu}=\mu/T$ .  

It is important to mention that the Seebeck coefficient of single species is independent of the relaxation time. We might mention here that the 
Seebeck coefficient has been estimated in condensed matter systems in the Drude limit. In this limit the Seebeck coefficient is
also independent of relaxation time as is shown explicitly in Ref.\cite{fujita}. 
It is also  straightforward to see from the definition
that the Seebeck coefficient is dimensionless. The Seebeck coefficient is a measure
of the fact that how efficiently any material can convert the temperature gradient to an electric current.
Material with higher Seebeck coefficient can more efficiently convert the temperature gradient into electrical current.
For low energy condensed matter system like metal e.g. copper ($Cu$), value of the Seebeck
coefficient is $S=1.05\mu$eV/K=10$^{-2}$, on the other hand for Semiconductors  $S=0.4$meV/K=4. For the details
about the Seebeck coefficient for conductors as well as semiconductors, see Ref.\cite{thermoelectrics}.
To get a feeling for the numerical values, in the present context, let us consider a single massless species and ignore the quantum statistics. In that case, from
Eq.\eqref{ssingle}, setting $m=0$, we have $S=(3/e-\mu/(eT))$ which for $\mu=0$ become $S\sim 9$.

\begin{figure}[!htp]
\begin{center}
\includegraphics[width=0.45\textwidth]{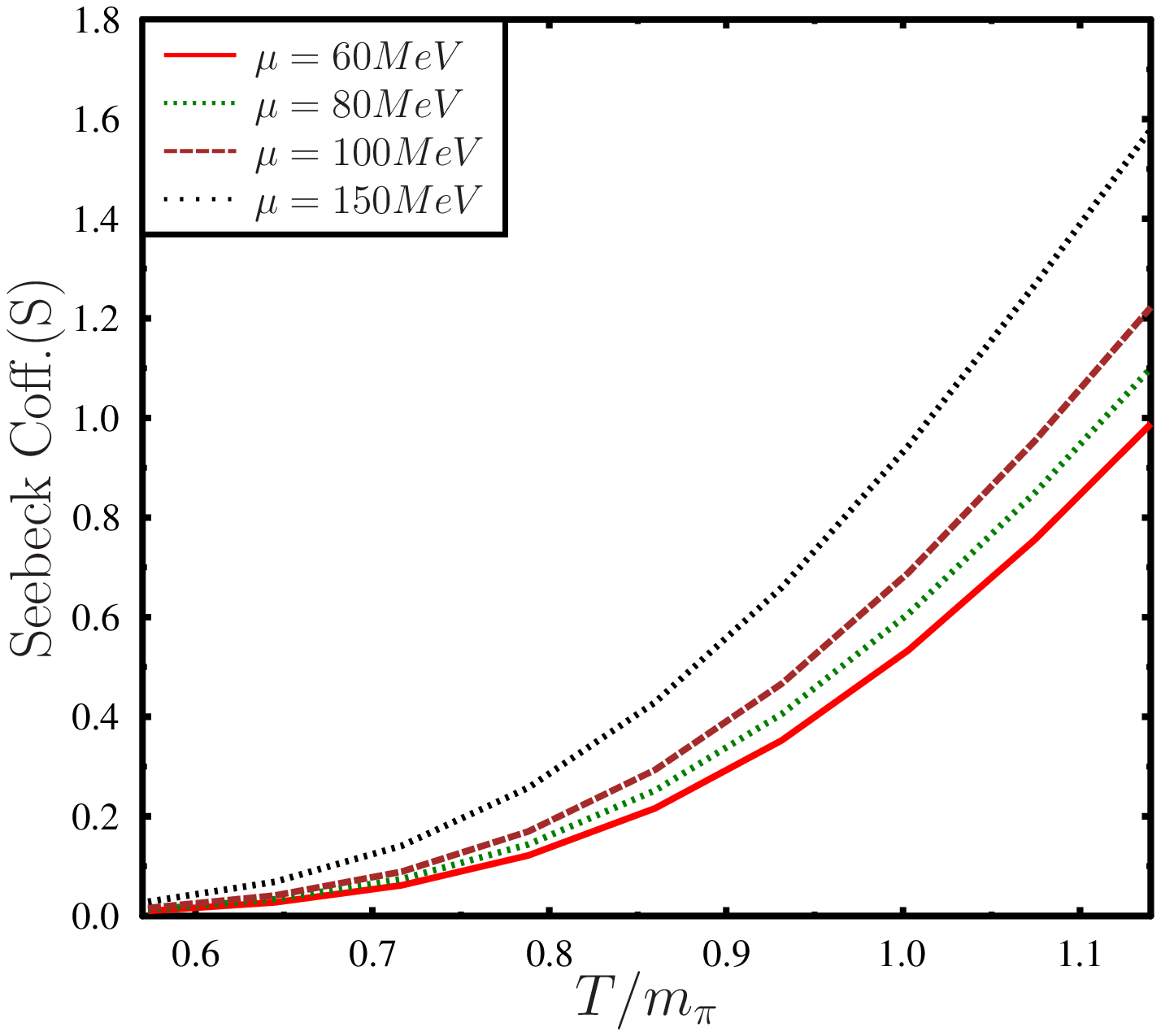}
\includegraphics[width=0.45\textwidth]{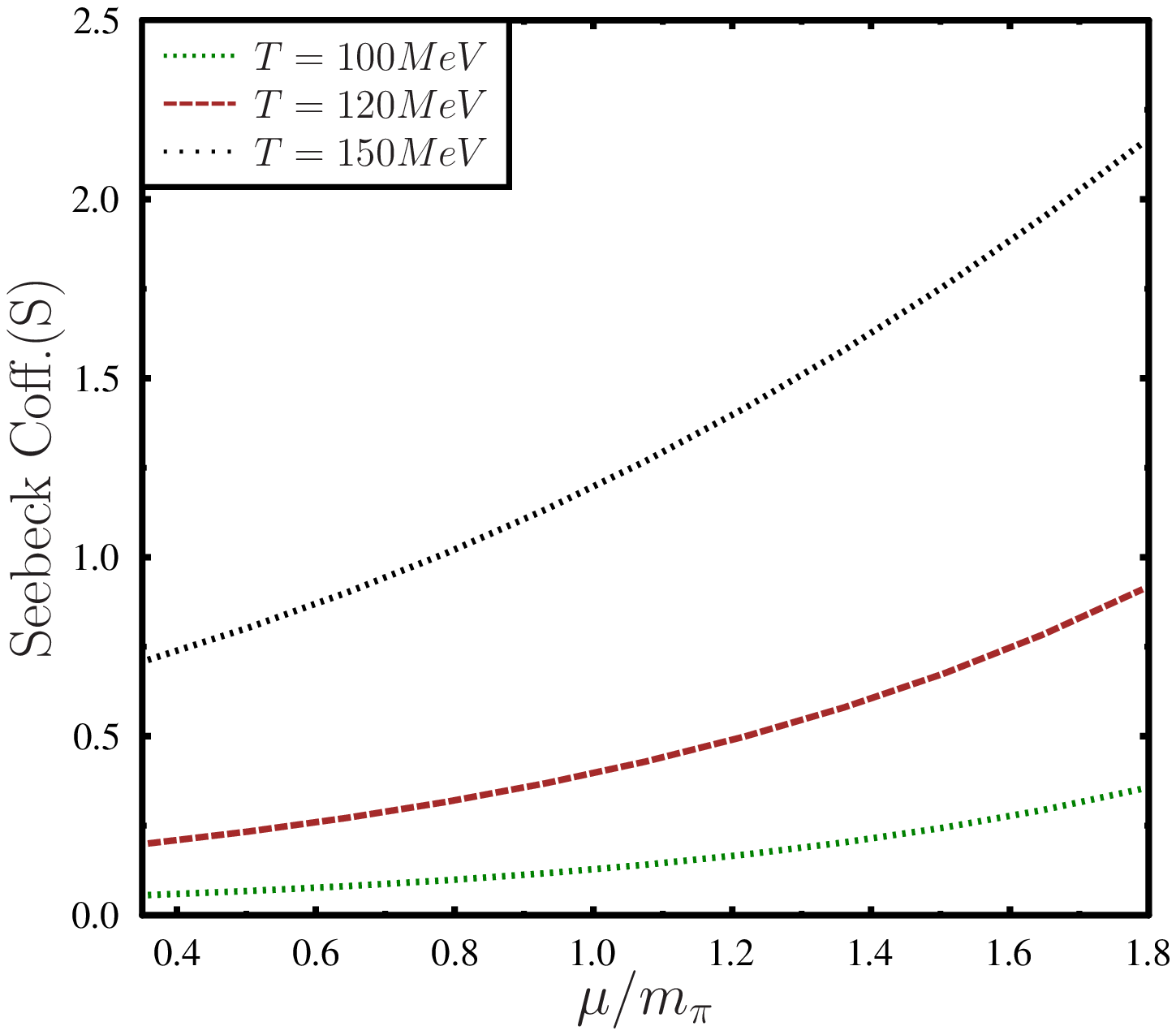}
\caption{Behavior of Seebeck coefficient (S) of hadron resonance gas as a function of temperature and baryon chemical potential.
We have used temperature range 80 MeV to 160 MeV, because degrees of freedom of HRG model are hadrons and resonances.
We have also taken the range of baryon chemical potential from 60 MeV to 250 MeV. In the left plot we have shown the variation 
of Seebeck coefficient of hadron resonance gas with temperature for different values of chemical potential. In the right plot 
we have shown the variation of Seebeck coefficient with baryon chemical potential for various temperature. 
In this calculation we have taken into account all the hadrons and resonances having mass up to 2.25 GeV.} 
\label{fig1}
 \end{center}
 \end{figure}
 
 We next show the variation of the total Seebeck coefficient for the hadronic medium of Eq.\eqref{eq38} with temperature (T) and 
 baryon chemical potential ($\mu$) in Fig.\eqref{fig1}. Let us note that the individual Seebeck coefficient $S^{(i)}$ as defined in Eq.\eqref{eq38} is independent of the corresponding relaxation time as it
 cancels out from the numerator and denominator. On the other hand the total Seebeck coefficient of the system is dependent on the relaxation time of
 individual hadrons through the $\mathcal{L}_{11}^{(i)}$s as in may be observed in Eq.\eqref{eq38}. The behavior of the 
 Seebeck coefficient as a function of baryon chemical potential can be understood from  Eq.\eqref{eq38} and Eq.\eqref{eq23}. Seebeck coefficient
 of the particle and the associated antiparticle is same but opposite in sign due to the explicit presence of the electric charge in $S^{(i)}$. Thus
 in the 
 numerator of the Eq.\eqref{eq38} mesons do not contribute.
 Hence only the baryons contribute to the in the numerator of the Eq.\eqref{eq38}. The mesons contributes in the denominator of the Eq.\eqref{eq38}.
 This is because in the denominator particles and antiparticles do not cancels out. In the denominator of the Eq.\eqref{eq38} mesons also take part because 
 in this case contribution from the particle and anti particle does not cancel out.
 
\begin{figure}[!htp]
\begin{center}
\includegraphics[width=0.45\textwidth]{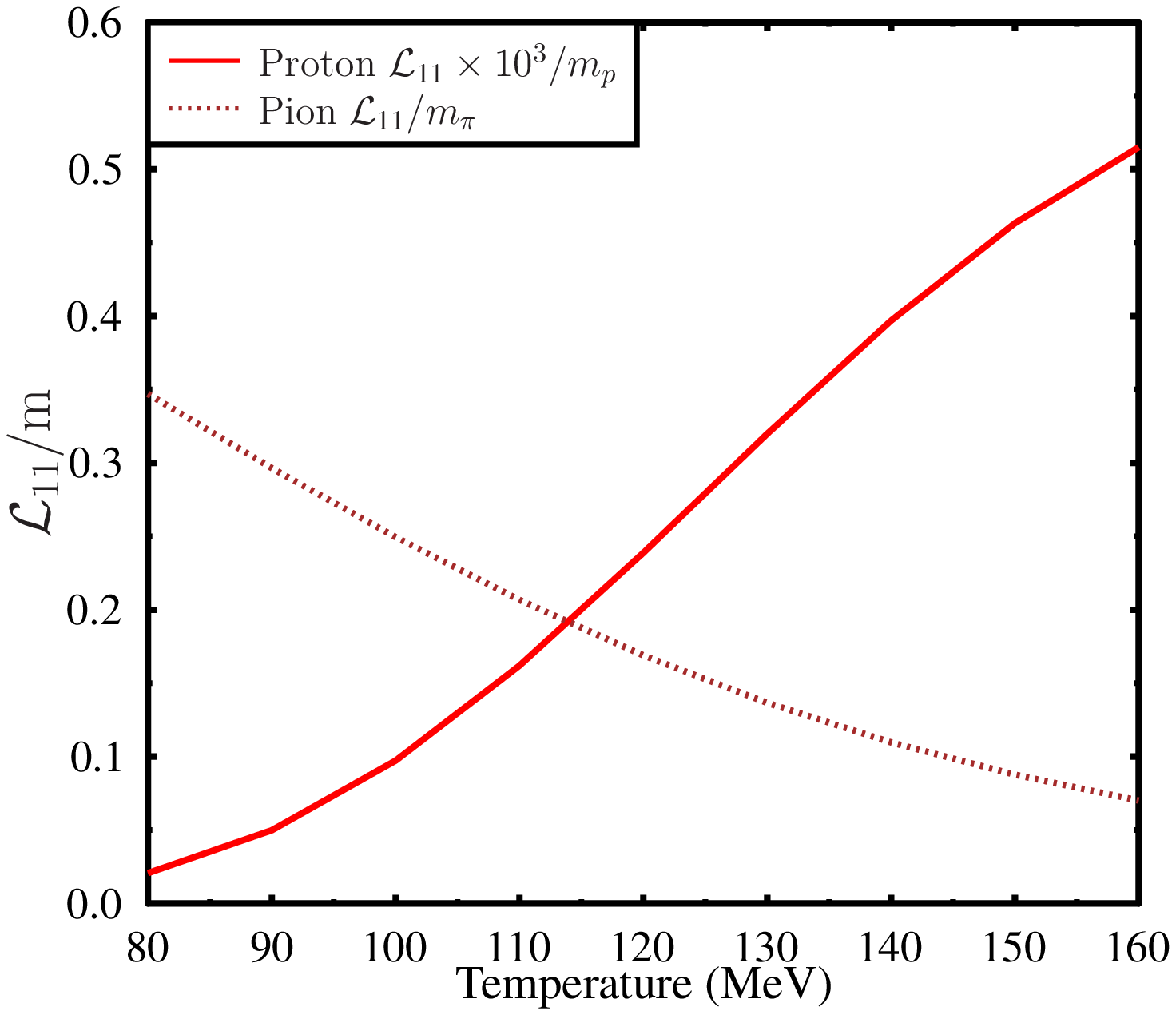}
\includegraphics[width=0.45\textwidth]{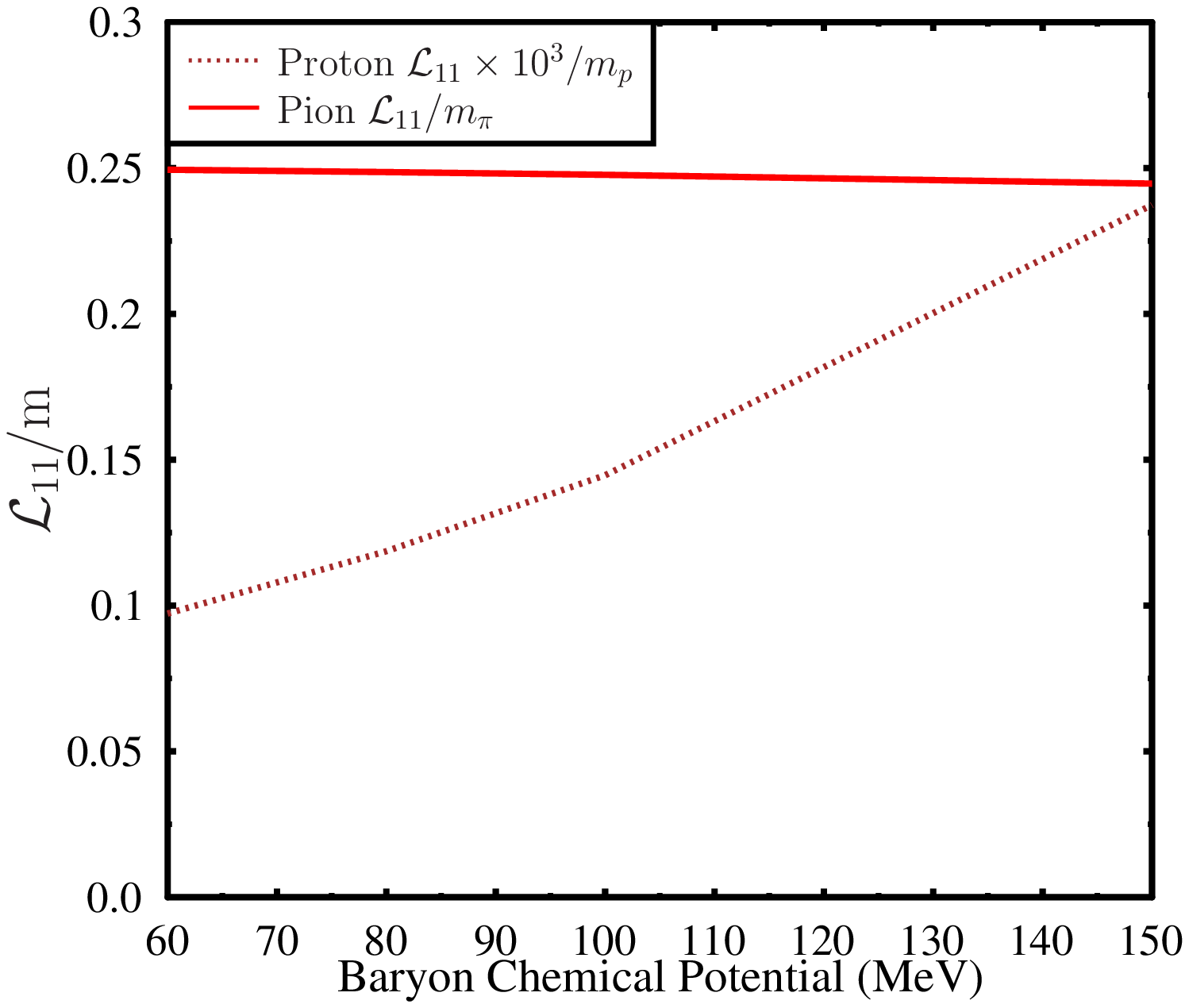}
\caption{\textit{Left plot}: Variation of  $\mathcal{L}_{11}$ of proton and pion with temperature for $\mu=60$ MeV has been shown for 
comparison. It is important to note that $\mathcal{L}_{11}$ of pion is very large with respect to the $\mathcal{L}_{11}$ of proton.
\textit{Right plot}: Variation of $\mathcal{L}_{11}$ of proton and pion with baryon chemical potential $\mu$ has been shown for T=100 MeV.
Dependence of Pion $\mathcal{L}_{11}$ on $\mu$ is due to the fact that relaxation time of pion depends weakly on the baryon chemical potential.} 
\label{fignew1}
 \end{center}
 \end{figure}

\begin{figure}[!htp]
\begin{center}
\includegraphics[width=0.48\textwidth]{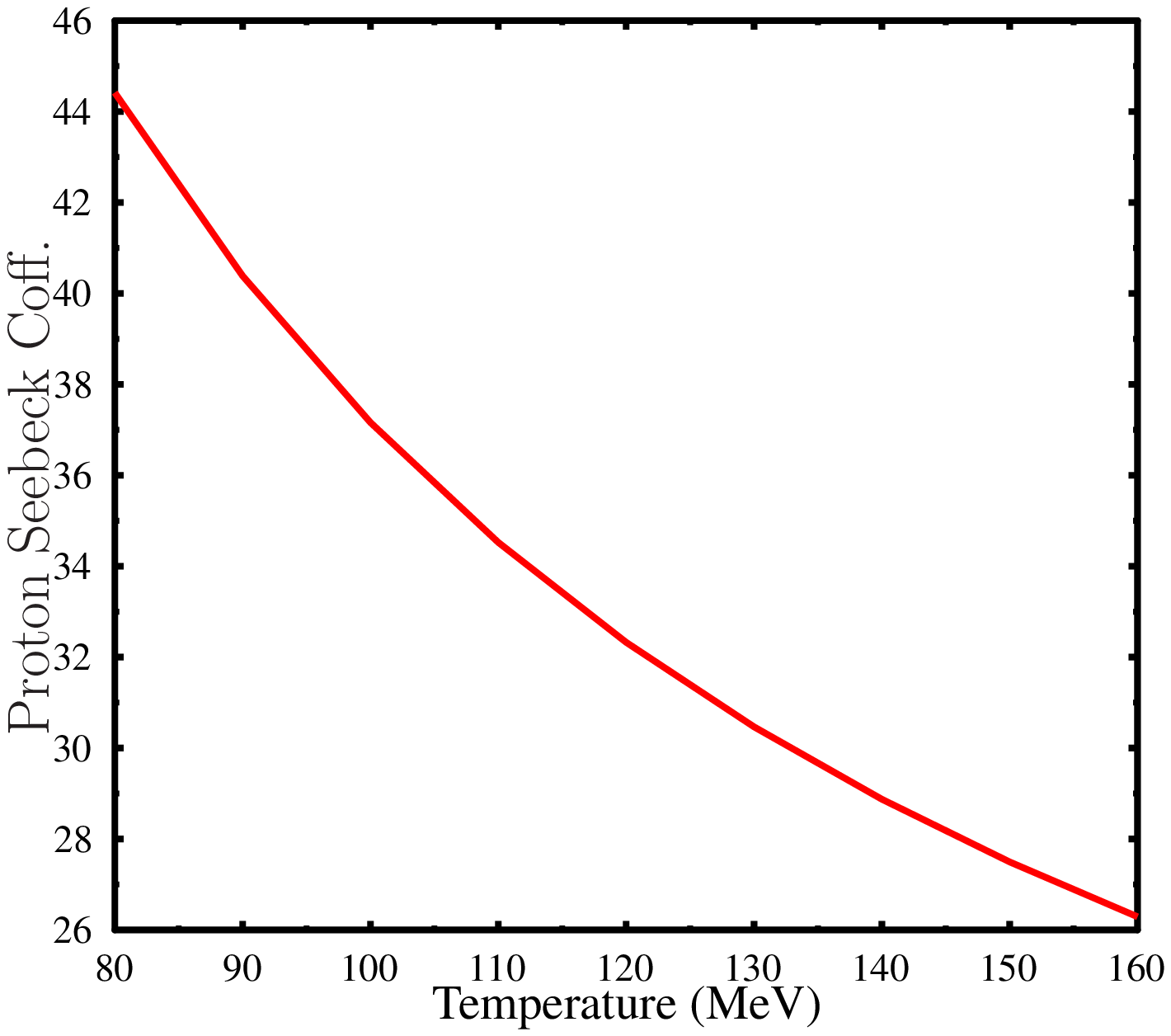}
\includegraphics[width=0.48\textwidth]{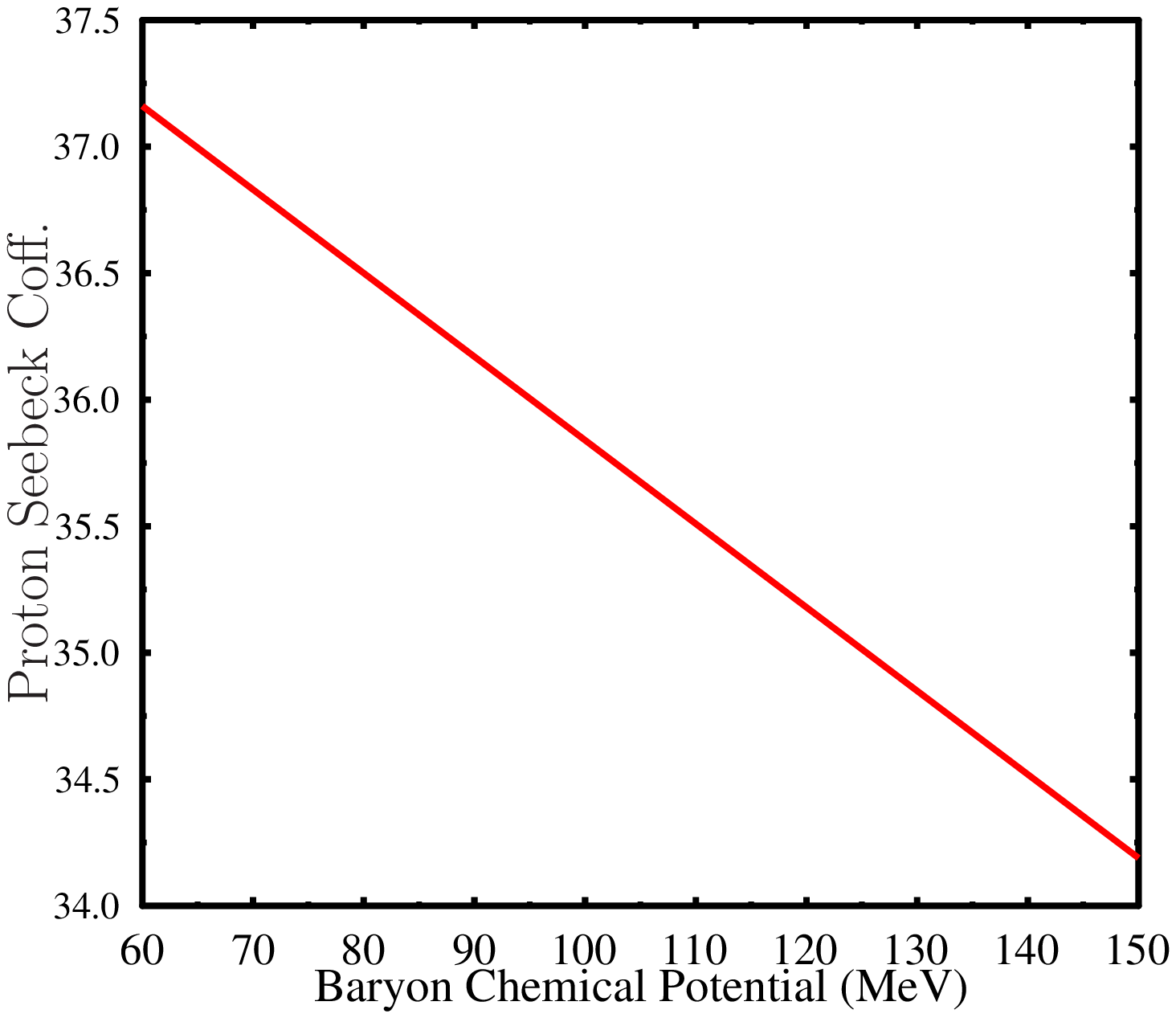}
\caption{\textit{Left Plot}: Variation of proton Seebeck coefficient as a function of temperature at $\mu=60$MeV.
This plot shows Seebeck coeff. of proton decreases with temperature for a given chemical potential. 
\textit{Right Plot}: Variation of proton Seebeck coefficient as function of baryon chemical potential at temperature $T=100 MeV$. 
This plot shows Seebeck coff. of proton decreases
with baryon chemical potential for a given temperature.} 
\label{fignew2}
\end{center}
\end{figure}

 To understand the behavior of the Seebeck coefficient with baryon chemical potential, let us first note that for the temperature ($T$), baryon
 chemical potential ($\mu$) range considered here the dominant contribution to the Seebeck coefficient arises from protons. The contribution from other 
 higher mass baryons are thermally suppressed. Behavior of proton Seebeck ($S^{(p)}$) shown in Fig.\eqref{fignew2},
 which decreases linearly with chemical potential
 ($\mu$) as may be obvious from Eq.\eqref{ssingle} . However the quantity $\mathcal{L}_{11}^{(p)}$ for proton increases with
 chemical potential ($\mu$) (Fig.\eqref{fignew1}). This increasing behavior of  
 $\mathcal{L}_{11}^{(p)}$ with $\mu$ is rather fast enough to make the product $S^{(p)} \mathcal{L}_{11}^{(p)}$ increasing with $\mu$.
 This make the numerator
 for the total Seebeck coefficient in Eq.\eqref{eq38} increases with chemical potential.
 This apart, for the denominator in Eq.\eqref{eq38} the dominant contribution arises for the pions.
 $\mathcal{L}_{11}^{(pion)}$ decreases with chemical potential as may be seen in right panel of Fig.\eqref{fignew1}. This
 decrease is due to the decrease of relaxation time for pions with increase in baryon chemical potential,
 through the hard sphere scattering. Taken together this explains the behavior of the total Seebeck coefficient 
 with baryon chemical potential ($\mu$), as be seen in Fig.\eqref{fig1}.
 
In a similar way one can understand the temperature dependence of Seebeck coefficient for the system of hadron resonance gas from the 
behavior of proton Seeback coefficient which is dominant in the sum given in Eq.\eqref{eq38} for the total Seebeck coefficient.
Note that as mentioned earlier in the 
sum Eq.\eqref{eq38} the mesons Seebeck coefficients do not contribute when the sum over the charged mesons are taken. With increasing 
temperature for a fixed chemical potential $\mathcal{L}_{11}^{(p)}$ of proton increases faster than the slow decreases of its Seebeck coefficient
as may be seen in Fig.\eqref{fignew1} and Fig.\eqref{fignew2} , making the numerator of Eq.\eqref{eq38} increases with $T$ for fixed $\mu$. This apart $\mathcal{L}_{11}^{(pion)}$
of pion also decreases with temperature for fixed baryon chemical potential as seen in the left panel of Fig.\eqref{fignew1}.
This leads to the increasing behavior of Seebeck coefficient as a function of temperature for a given baryon chemical potential, as seen in
Fig.\eqref{fig1}.

It might be relevant here to note that the thermoelectric current so produced due to the temperature gradient can generate a 
magnetic field in the heavy ion collision experiments. One can estimate an order of magnitude of the magnetic
field produced due to this thermoelectric current. The magnitude of electric current 
density produced by the temperature gradient can be expressed as,

\begin{align}
 j=\sigma_{el} S \nabla T,
\end{align}

It may be seen in Fig.\eqref{fig1} the order of magnitude of the total Seebeck coefficient of the hadron resonance gas can be of $\mathcal{O}$(1) for
temperature $T=120$ MeV.
The electrical conductivity of hot pion gas can be taken to be of the order of $\sigma_{el}/T \sim 0.01$ \cite{MoritzGreif}.
Therefore for the temperature of the order of 100 MeV, electrical conductivity is of $\mathcal{O}(1)$ MeV. If we assume that the system
size is about 20 fm and the temperature
difference between central and peripheral region to be of the order of 100 MeV then the temperature gradient is of the order of $10^3$ MeV$^2$.
This leads to electrical current density 
is of the order of $10^3$ MeV$^3$. If we take the cross-sectional area to be (20 fm)$^2$, then the electrical current is $\sim$ 10 MeV.
Magnetic field 
generated by the current $I$ can be given as, $B = \frac{I}{2\pi r}$. So, for $r\sim 20$ fm,
the magnetic field is $\sim$ 16 MeV$^2 \sim 10^{-3} m_{\pi}^2$. However we must note that this is a very crude approximation for the 
magnitude of the field generated as the heavy ion collisions are highly dynamic in nature. Therefore the magnetic field generated can only be 
a transient one. Further a more realistic estimation will require a full dynamical calculation using possibly a transport simulation.
Such a magnetic field is not the remnant of the
initial magnetic field produced in heavy ion collisions, rather the source of this magnetic field is the current produced due
to the temperature gradient in a baryon rich plasma.

\section{Conclusion}
In this work, we have attempted to study the thermoelectric effect of a thermalized hadronic medium with a temperature gradient.
We have estimated the corresponding Seebeck coefficient of hot hadronic matter within hadron resonance gas model (HRG). 
Thermoelectric effect necessarily requires a temperature
gradient which is achievable in heavy ion collision experiment due to the temperature difference in the central and peripheral part of the fireball
produced in these collisions. One of the important outcomes of this calculation is that, for a baryon free plasma, contributions in total Seebeck
coefficient due to the mesonic degrees of freedom cancel out.
This happens because of the fact that each meson particle comes with its antiparticle with opposite charge leading to cancellation of the corresponding
Seebeck coefficient for the charged mesons. However, in a baryon rich plasma, the contributions to the total Seebeck coefficient due to the baryons
do not cancel out. Total Seebeck coefficient of thermalized 
hadron resonance gas increases with increasing temperature for fixed baryon chemical potential and increases with baryon chemical potential for fixed
temperature. It is important to note that the formalism we are using 
is a non-relativistic one. It will be important to study the thermoelectric effect in a relativistic formalism, particularly for QGP medium.
Electrical current produced due to the temperature gradient can be a source of magnetic field. According to our crude estimate the strength of the 
transient magnetic field so generated can be $\sim 10^{-3} m_{\pi}^2$. However, the magnetic field so produced through thermoelectric effect crucially
depends upon the temperature
gradient, the thermal profile of electrical conductivity and the Seebeck coefficient. In this work, we have discussed the formalism to calculate
the thermoelectric
coefficient in the case of spatially uniform baryon chemical potential. In a more general scenario, there can be a spatial variation of 
baryon chemical potential. In that case there can be current generation driven by the chemical potential gradient. Although we cannot
make any comment at present on the current generation due to both the temperature gradient and chemical potential gradient, it will never the less
be interesting
to study these effects. In particular  
current generation due to a chemical potential gradient might be interesting for high baryon density matter. 

\section*{Acknowledgements}
The idea discussed in the present investigation arose during a visit of one of the author's (HM)
to the research group of Prof. Ajit M. Srivastava at Institute of Physics Bhubaneswar.
The authors would like to thank Ajit. M. Srivastava for suggesting the problem and members
of his research group for subsequent extensive discussions. The authors would also like
to thank Sabyasachi Ghosh, Abhishek Atreya, Aman Abhishek, Chowdhury Aminul Islam, Rajarshi Ray
for many discussions during working group activities at WHEPP 2017, IISER Bhopal. 
We thank Guruprasad Kadam for useful comments on the manuscript.

\end{document}